\documentclass[%
 reprint,
 amsmath,amssymb,
 aps,
prl,
]{revtex4-1}

\usepackage{graphicx}% Include figure files
\usepackage{dcolumn}% Align table columns on decimal point
\usepackage{bm}% bold math
\usepackage{amssymb} 
\usepackage{braket}
\usepackage{mathcomp}
\usepackage{amsmath}

\begin{document}

\preprint{APS/123-QED}

\title{Quantum entanglement, two-sided spacetimes and the thermodynamic arrow of time}% Force line breaks with \\

\author{Ovidiu Racorean}
 \email{ovidiu.racorean@mfinante.gov.ro}
\affiliation{%
 General Direction of Information Technology, Bucharest, Romania\\
}%

\date{\today}% It is always \today, today,

\begin{abstract}
We investigate the emergence of thermodynamic arrow of time in the context of AdS/CFT correspondence. We show that, on the CFT side, if the two copies of the field theory are not initially correlated the entropy can only increase such that a definite orientation of the thermodynamic arrow of time is imposed. Conversely, in a high-correlation environment, the entropy can either increase or decrease, such that there is no opportunity for the dominance of one direction of time over the other. On the gravity side, we construct the structure of geometric dual by considering the notion of spacetime sidedness and time-orientability. Accordingly, we conjecture that the entanglement of the CFTs in the thermofield double state, impose the connection of the two sides of the spacetime forming a one sided spacetime. In addition, disentangling the degrees of freedom of the two CFTs, results in disconnecting the two sides of the spacetime. In essence, the maximal entanglement between the two copies of CFT builds, in the geometric dual, a connection between the two sides of spacetime.
\begin{description}
\item[PACS numbers]
11.25.Tq, 03.65.Ud, 03.67.-a, 04.20.Gz

\end{description}
\end{abstract}

                            
\maketitle

%\tableofcontents

\section{\label{sec:level1}Introduction %\protect\\
%The line break was forced \lowercase{via} \textbackslash\textbackslash}
}
Recent work on the AdS/CFT correspondence has unveiled remarkable connections between quantum information theory and gravity. In the context of AdS/CFT duality, the structure of quantum entanglement in the CFTs state is intimately related to the geometrical structure of the dual spacetime \cite{mal}, \cite{sus}.  Furthermore, it has been argued in \cite{mal}, \cite{sus}, \cite{raa}, \cite{mark}, that the emergence of classically connected spacetimes is the consequence of the entanglement structure of the underlying quantum mechanical degrees of freedom.

The initial conditions of the correlations between the CFTs influence the dual geometry of spacetime. Therefore, the initial product state of the CFTs (no initial correlation between the two quantum subsystems) is dual to two disconnected AdS spacetimes, while classical connectivity between the spacetime pair arises as result of initial entangled states. In the latter case, it turns out that varying (increasing or decreasing) the degree of entanglement between the two copies of field theory, thus the entanglement entropy, results in varying the degree of connectivity in the dual spacetime \cite{raa}, \cite{mark}.

The same argument of initial conditions between two quantum systems has been reiterated in recent debate \cite{par}, \cite{seth}, \cite{jen}, \cite{jev}, \cite{mac}, \cite{parr}, \cite{ved} on the emergence of thermodynamic arrow of time. Consequently, it has been reported \cite{mic} that the lack of initial correlations between the subsystems results in the emergence of a preferred direction of the arrow of time, as is reflected by the standard second law. The presence of initial quantum correlations in the local thermal states, on the other hand, can result in reversing the arrow of time. Specifically, lowering the initially high correlations (high degree of entanglement) among the two quantum subsystems, thereby lowering their individual entropies, results in reversing the thermodynamic arrow.

In this Letter we investigate in the landscape of AdS/CFT duality, the emergence of the thermodynamic arrow of time on the field theory side and how is reflected in the geometric structure of the dual spacetime. Consequently, we start, on the CFT side, in a low-correlation environment with the two copies of CFT in a product state. In this case, the mutual information and consequently the entropy can only increase, such that we are in the situation considered in the formulation of the second law of thermodynamics \cite{seth}, \cite{cas}, \cite{fey}. Specifically, the arrow of time is oriented toward the standard thermodynamic arrow.  Additionally, the time is oriented in opposite directions in the individual CFT’s. 

In contrast, in a high-correlation environment with the two CFT copies entangled in thermofield double state allow the individual entropies to change in both directions. In this scenario, the initial entangled state between the two copies of field theory permits the individual entropies to increase or decrease, depending on the initial mutual information. Under these circumstances, we argue that on the composite system there is no preferred orientation of time, there is no opportunity for the dominance of one direction of time over the other \cite{par}, \cite{mic}.

On the gravity side, we argue that in the presence of initial correlations, since the composite quantum system exhibits no definite direction for the arrow of time, the geometric structure of the spacetime is consistent with that of a time-unorieted, thus a one-sided spacetime. In this scenario the two sides of the spacetime are connected in the presence of high correlations forming an unoriented, one-sided spacetime. Moreover, when decrease the degree of entanglement between the dual CFTs, the two dual spacetime regions disconnect from each other, ending up with a two sides of spacetime. The reader may find it useful to recall at this point the notions of spacetime orientability and two-sidedness in \cite{haw}, \cite{hil}, \cite{had}.

In the absence of correlations, the pair of spacetimes is disconnected and a time-oriented structure of the composite spacetime is imposed. Consequently, the spacetime is consistent with an oriented, two-sided spacetime.

Based on these arguments, we conjecture that the geometric structure of spacetime is consistent with a one-sided spacetime in the presence of initial high degree of entanglement in the dual composite quantum system, while reducing the entanglement results in disconnecting the two sides of spacetime, eventually ending up in a two-sided spacetime.

The results in this Letter suggest that, in essence, the maximal entanglement between the two copies of CFT builds, in the geometric dual, a connection between the two sides of spacetime.

\section{\label{sec:level2}Initially uncorrelated states}

We start, as usual, considering two non-interacting copies of CFT on sphere $S^d$ (x time). The two CFTs are equivalent to two subsystems noted $L$ and $R$, such that we can decompose Hilbert space $H_{LR}$ of the composite system as $H_{LR}=H_L\otimes H_R$ . We consider that the two subsystems, L and R, respectively are initially uncorrelated to begin with. Thus, in this case, since there is no entanglement between the two CFTs components, we have two completely separate physical systems which do not interact, such that the joint state of the system is the product state: 

\begin{equation}
\rho_{LR}=\rho_L\otimes \rho_R.
\end{equation}
  				                              
Here, the quantities $\rho_L$, $\rho_R$ and $\rho_{LR}$ are the density matrix describing the states of the left, right ant the composite system, respectively. It is important to note here that  $\rho_L$, $\rho_R$ correspond to the thermal state of the left and right copies of field theory.

On the gravity side of the AdS/CFT correspondence, the interpretation of this initially uncorrelated state is straightforward \cite{mal}, \cite{raa}. The two separate physical systems determined by the density matrix $\rho_L$ and $\rho_R$, correspond in the dual description, to two separate asymptotically AdS spacetime. We can emphasize here that the joint state in Eq.(1) is dual to a disconnected pair of spacetimes.

Let us return to the initial product state of the two copies of field theory in Eq.(1) and consider employ here as a natural measure of the correlations between the two subsystems, $L$ and $R$, the mutual information:

\begin{equation}
I(\rho_{LR} )= S(\rho_L )+S(\rho_R )-S(\rho_{LR} ),
\end{equation}
		  		                                     	 
where $S(\rho_L )$, $S(\rho_R )$ and $S(\rho_{LR})$ are the entropies of the left, right and the composite system, respectively. We have considered here, as usual, the von Neumann entropy, $S(\rho)=-Tr(\rho log\rho)$, of the density matrix , to define the entropies. 

Since we are now in a low-entropy environment, due to the product state in Eq.(1), we know that $I(\rho_{LR})= 0$, which reduces the Eq.(2) to: 

\begin{equation}
S(\rho_{LR})=S(\rho_L)+S(\rho_R).
\end{equation}
                                                         
As result, the mutual information and consequently the entropy of the composite system can only increase. To see this, let us now consider entangling some of the degrees of freedom of the individual components, evolving in this way the entropy of the joint state from the initial,

\begin{equation}
S_i(\rho_{LR})=S_i(\rho_L)+S_i(\rho_R),
\end{equation}
 
to the final entropy:  

\begin{equation}
S_f(\rho_{LR})\leq S_f(\rho_L)+S_f(\rho_R).
\end{equation}
 
The initial and final states of the composite system, $\rho_{LR}^i$ and $\rho_{LR}^f$ respectively, are related unitarily, thus we have, $S_f (\rho_{LR})=S_i (\rho_{LR})$. With the help of Eq.(4) and Eq.(5) we find that:

\begin{equation}
\Delta S(\rho_L )+\Delta S(\rho_R )\geq0.
\end{equation}
								
This result is consistent with the second law of thermodynamics which states that the entropy can only increase. Generally, the total entropy can only increase since the mutual information is zero. In this case of the absence of initial correlations between the two CFT’s, the entropy can only increase, in accordance with the formulation of the second law of thermodynamics, such that we have a precise direction of the arrow of time oriented in the sense of increasing entropy. Specifically, the flow of time is directed toward the standard thermodynamic arrow.

We have now a definite orientation of time for the joint state of the CFTs, such that on the gravity side, we can interpret the product state of the CFTs as dual to a time-oriented spacetime with two disconnected components. Since the two components of the global spacetime are disconnected we may ask at this point how is the arrow of time oriented in these individual regions of the spacetime. This question is legitimate since we are working in the framework of AdS/CFT duality such that we may consider that each of the two independent copies of the field theory have its own time \cite{mal}. Thus, to find the thermodynamic arrow on the two disconnected spacetimes let us now focus on the changes in individual entropies of CFT subsystems. 

We can find indications on the changes in the individual entropies in Eq.(6). Since the entropy of the composite system must be positive, a decrease of the individual entropy in one subsystem must be accompanied by an increase of entropy in the other. This statement find supported in \cite{seth} where it has been suggested that decreasing the entropy by $\Delta S$ in the left subsystem,  $S_L^f=S_L^i-\Delta S$ is followed by an increase of entropy in the right subsystem by at least  $\Delta S$:

\begin{equation}
\Delta S(\rho_R )\geq \Delta S.
\end{equation}
											
This result indicates that, in essence the evolution of the individual entropies is exactly opposed in the two subsystems, $L$ and $R$; while the entropy decreases in one subsystem, it increases in the other. Consequently, from the point of view of the second law, while the time is oriented in the standard thermodynamic direction in one system, it is oriented in reverse in the other system.

Now, we may conclude that on the gravity side, the structure of the geometric dual is that of a spacetime with a preferred orientation of time in the sense of standard thermodynamic time. We can argue here, that the lack of initial correlations between CFTs ensures a time-orientation in the dual spacetime. Furthermore, since the arrow of time is oriented in opposition in the two individual CFTs, in the dual asymptotically AdS components the time is oriented in opposite direction. On the gravity side the initial product state of the CFTs is interpreted as a time-oriented spacetime with two disconnected components on which the time is oriented in opposite direction.

\section{\label{sec:level3}Initially entangled states}

We consider now that the two copies of the field theory are high-correlated to begin with. In this scenario the situation changes dramatically. To be more specific, we further consider that the two copies of CFT are initially entangled in the thermofield double state. The joint state of the two subsystems thus can be represented, in the context of AdS/CFT correspondence, as:

\begin{equation}
\rho_{LR}=\ket{\Psi_\beta} \bra{\Psi_\beta},
\end{equation}
										
with $\ket{\Psi_\beta}$  defined as the thermofield double state,

\begin{equation}
\ket{\Psi_{\beta}} = \frac{1}{\sqrt{Z_{\beta}}}\sum_ne^\frac{-\beta E_n}{2}{\ket{E_n}}_L{\ket{E_n}}_R,
\end{equation}

where  $〖\ket{E_n}〗_L (〖\ket{E_n}〗_R)$ is the $n$-th energy eigenvector for the subsystem $L (R)$ , $\beta$ is the inverse temperature and $(Z_\beta)^-\frac{1}{2}$ is a normalization constant. Since Eq.(9) is essentially the Schmidt decomposition of $\ket{\Psi_\beta}$ , we may remark here that the two subsystems are initially entangled in a pure state. Note however that the individual states of the two so constructed subsystems are the thermal states. We can see this by tracing over the Hilbert space of the other, $\rho_L=〖Tr〗_R \rho_{LR}$.  

The thermofield double state in Eq.(9) is understood \cite{mal}, \cite{raa} on the gravity side as a spacetime having two asymptotically AdS spacetimes components classically connected.

We further expand on the Van Raamsdonk’s idea \cite{raa} on disentangling some of the degrees of freedom between the two CFTs. We ask what happened with the individual entropies in the two subsystems once we decrease de degree of entanglement in the joint system. 

We start now in a high-correlation environment. As it was shown in \cite{raa}, disentangling some of the degrees of freedom is consistent with the decrease of mutual information. Consequently, the entropy also decreases. To see this clearly, we consider decreasing the entropy in the left CFT by $\Delta S$ such that the final individual entropy is  $S_L^f=S_L^i-\Delta S$. We know from \cite{raa} that in this case the change in entropy in the right CFT must be: 

\begin{equation}
\Delta S(\rho_R )\geq\Delta S-I_{LR}^i,
\end{equation}
										
where $I_{LR}^i$ is the initial mutual information. 

In stark contrast to the uncorrelated case, discussed earlier, here, the presence of initial correlation $I_{LR}^i$ alters the outcome of the inequality. While a low initial mutual information allow the entropy of the right CFT to increase, an initial high-correlation environment force the entropy in the right CFT to also decrease. In this letter case both individual entropies decrease. Moreover, we emphasize here that considering the initial state as a pure state of the two copies of the field theory in the thermofield double in Eq.(9) ensures that $\rho_L$ and $\rho_R$ are isospectral so that $S(\rho_L )=S(\rho_R )$.

The initial entanglement of the composite system forces the individual entropies $S(\rho_L)$ and $S(\rho_R)$ to move in the same direction, such that:

\begin{equation}
\Delta S(\rho_L)=\Delta S(\rho_R ),
\end{equation}
										
at all times.
In addition, we can say that the initial pure state, reflected in the initial maximal entropy, imply that the individual entropies can only decrease, such that $\Delta S(\rho_L)=\Delta S(\rho_R)<0$.
In this case, Eq.(6) is reversed as:

\begin{equation}
\Delta S(\rho_L )+\Delta S(\rho_R )\leq 0,
\end{equation}
  					
a result which suggests that the entropy can only decrease. Since the  the thermodynamic arrow is reversed. From the perspective of the second law of thermodynamics \cite{cas}, \cite{fey} the decrease of individual entropies implies a reversal of the arrow of time, since the natural orientation of the arrow of time, from past to future, is in the sense of increasing entropy.

We see that changes in the individual entropies in the scenario of initial high-correlation environment allow both orientations of the thermodynamic arrow, such that there is no opportunity for the dominance of one direction of time over the other \cite{par}, \cite{mic}.  

On the gravity side, we know \cite{raa} that the geometric structure of the dual spacetime is that of a spacetime with two asymptotically AdS components classically connected. Moreover, considering the changes in the individual entropies, from the perspective of AdS/CFT correspondence, the dual spacetime of the composite system is a spacetime with no preferred orientation of time. The initial high correlations of the two CFT’s ensure that in the dual spacetime the arrow of time may be oriented normal, from the past to the future or in reverse, from the future to the past. In this context, since there is no preferred orientation of time on the spacetime, we may conjecture that the gravity dual is a time-unoriented spacetime with two connected components. 

\section{\label{sec:level4}Orientability and sidedness of the spacetime}

Let us now advocate the disentangling idea further and consider decreasing the entanglement between the CFTs to zero. Our intention is to see what happen with the thermodynamic arrow in the process.

On the CFT side, we start with the two copies of the field theory entangled in the thermofield double state and gradually disentangle the degrees of freedom to zero. As result, decreasing the entanglement to zero between the two CFTs components, we remain in the end with two completely separate physical systems which do not interact such that the joint state of the system is $\rho_{LR}=\rho_L\otimes \rho_R$. Now, from the arrow of time perspective, we start with a state with no preferred orientation of the arrow of time and decreasing the entanglement to zero the end result is a joint state that has a definite orientation of time in the sense of standard thermodynamic arrow. 

Let us now evaluate how decreasing entanglement to zero is reflected in the dual geometric structure of spacetime. On the gravity side, we start with a spacetime with two asymptotically AdS components classically connected as the dual to the entangled state of the CFTs. Using Ryu-Takayanagi proposal, van Raamsdonk has argued in \cite{raa}, \cite{ryu} that decreasing the entanglement is equivalent in the dual spacetime with a decrease of the minimal surface which separates the two components of the spacetime. Accordingly, the two regions of dual spacetime, initially connected, gradually disconnects from each other in direct proportionality to the decrease of entanglement. Thus, when entanglement is decreased to zero the two regions are pinching off and disconnect from each other. The end result is a spacetime with disconnected spatial regions.

In this scenario thus arises the question of what happens with the arrow of time on the structure of the geometric dual. More specifically, how the arrow of time changes when decreasing the entanglement to zero we pass from a spacetime with two components connected to a spacetime with disconnected components?

We argued that the spacetime with connected regions we start with is a spacetime that have no preferred orientation of time, hence a time-unoriented spacetime characterized by the lack of opportunity for the dominance of one direction of time over the other. From this point, we decreased the area separating the two components of the spacetime to zero.  As result the two components of the spacetime disconnect from each other ending up in a spacetime with a preferred orientation of time. In summary, we can argue that starting with a time-unoriented spacetime and decreasing the entanglement to zero we end up eventually with a time-oriented spacetime divided in two spacetime regions. Moreover, the arrows of time on the two disconnected components of the spacetime have opposite orientation.

This behavior characterizes the sidedness of spacetime. The logical arguments to support this statement are as follows. The end result of decreasing the entanglement is a time-oriented spacetime with two disconnected components. Moreover, time on these two spacetime components is oriented in opposite direction. At this stage we can identify the two disconnected spacetime regions with the two disconnected sides of a two-sided spacetime. The reader may find it useful to recall at this point the notions of orientability and sidedness of spacetime in \cite{haw}, \cite{hil}, \cite{had}.

On the other hand, initially, we have started with a time-unoriented spacetime with two connected components. Accordingly, we can consider that initially the two sides of the spacetime were connected, and gradually disconnect in direct proportionality to the decrease of entanglement. Since we have a time-unoriented and two connected sides of spacetime, we argue that the spacetime is consistent with a one-sided spacetime. 

Intuitively, we can think the geometric dual of disentangling the degrees of freedom as    cutting the one-sided spacetime over the minimal surface to produce a two-sided spacetime. The process is sketched in the Fig.1

\begin{figure}
\includegraphics[width=8.6cm]{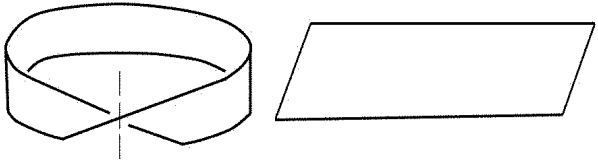}
\caption{\label{fig:fig1} Cutting the one-sided spacetime (left) over the interrupted line, seen here as the area separating the two spatial regions, results in a two-sided spacetime (right).}
\end{figure}

Based on these arguments we are lead to a surprising conclusion; the entanglement of two copies of the field theory can be identified, in the gravity dual, with a connection between the two-sides of the spacetime.

At this point, we can make a precise statement of what happen on the structure of geometric dual when we vary the degree of entanglement between the two individual quantum subsystems, $L$ and $R$. Consequently, starting in a high-correlation environment and gradually disentangling the degrees of freedom between the two CFTs, hence lowering the mutual information, results in disconnecting the two-sides of spacetime. As result, the initially one-sided spacetime, sustained by high-correlations of the CFTs, reduces to a final two-sided spacetime in the absence of correlations between the individual subsystems.

We argue that, in essence, the maximal entanglement between the two copies of CFT builds a connection between the two sides of spacetime, while disentangling the degrees of freedom is reflected on the geometric dual by disconnecting the two sides of the spacetime.

\section{\label{sec:level5}Conclusions}

We have discussed the emergence of the arrow of time in the AdS/CFT correspondence framework. We have shown in this Letter, that the direction of the thermodynamic arrow of time on the individual asymptotic spacetimes is dictated by the degree of entanglement on the dual CFTs. 

Accordingly, the lack of initial correlations between the two copies of CFT constrains the entropy of the composite system to only increase. As result, the thermodynamic arrow of time is oriented natural, according to the second law. On the other hand, the initial high-correlation present in the joint state of the CFTs ensures   In this case, we argue that the thermodynamic arrow may be oriented in both directions; there is no opportunity for the dominance of one direction of time over the other.

Based on these arguments, we have conjectured that the structure of the spacetime geometry is consistent with that of a time-unorionted spacetime with two connected components, in the presence of the initial high-correlations, whereas in the case of low initial correlations the geometry is that of a time-oriented spacetime with two disconnected components. Therefore, starting with initial high-correlation and disentangling the degrees of freedom is consistent, in the geometrical dual, with gradually disconnecting the two sides of spacetime ending up in a two-sided spacetime.

In essence we have argued that the maximal entanglement between the two copies of field theory builds a bridge connecting the two sides of spacetime.

\end{document}